\documentclass[amsmath,amssymb,showkeys]{revtex4}

\usepackage{graphicx}
\usepackage{dcolumn}
\usepackage{bm}
\usepackage{amssymb}
\usepackage{amsmath}
\usepackage{longtable}
\usepackage{topcapt}
\begin{document}


\title{Optimization and evaluation of a coarse-grained model of protein motion using X-ray crystal data}

\author{Dmitry A. Kondrashov$^*$, Qiang Cui$^\dagger$, and George N. Phillips Jr$^{\dagger,\ddagger}$}
\affiliation{%
$^*$Department of Biochemistry, \\ $^\dagger$ Department of Chemistry and Theoretical Chemistry Institute, \\
$\ddagger$ Departments of Biochemistry and Computer Science \\
University of Wisconsin - Madison, Madison, WI, USA
}%
 \email{phillips@biochem.wisc.edu}
 
\date{\today}

\begin{abstract}

Comments: RevTeX, 18 pages, with 4 figures \\
Subj-class: Biomolecules \\ \\

Simple coarse-grained models, such as the Gaussian Network Model, have been shown to capture some of the features of equilibrium protein dynamics.  We extend this model by using atomic contacts to define residue interactions and introducing more than one interaction parameter between residues.  We use B-factors from 98 ultra-high resolution X-ray crystal structures to optimize the interaction parameters.  The average correlation between GNM fluctuation predictions and the B-factors is 0.64 for the data set, consistent with a previous large-scale study.  By separating residue interactions into covalent and noncovalent, we achieve an average correlation of 0.74, and addition of ligands and cofactors further improves the correlation to 0.75.  However, further separating the noncovalent interactions into nonpolar, polar, and mixed yields no significant improvement.  The addition of simple chemical information results in better prediction quality without increasing the size of the coarse-grained model.

\end{abstract}


\maketitle

\newpage
\section*{Introduction}

Proteins reliably self-organize into specific shapes that are essential for their function. The coordinates that are reported as protein structures, however, are the average positions of an ensemble of fluctuating conformers that constitute the native state. It is becoming increasingly accepted that protein structures define specific types of motions that play important roles in protein function. The mechanism is rarely clear, however, owing in part to the difficulty of direct observation of protein motions. Crystals can be subjected to time-resolved experiments \cite{schotte2003-1944}, but the range of applications is limited to reactions that can be triggered by light or trapped by clever manipulations.  NMR spectroscopy is can be used to determine both the structure and the dynamics of proteins \cite{lindorff2005-128}, but it is limited both by the maximum size of protein structures and by the difficulty of discrimination of slowly or quickly exchanging dynamics \cite{palmer2001-204}. Mass spectrometry coupled with hydrogen/deuterium exchange and proteolysis has been used to determine changes in the relative solvent accessibility of amide hydrogens \cite{lanman2004-181}, and single-molecule experiments using optical trapping have resulted in spectacular observations of the motion of motor proteins \cite{abbondanzieri2005-460}. Overall, direct measurement of molecular motion remains laborious and limited.

Computational methods have been utilized for several decades to study the motion of proteins \cite{karplus2005-6679}, but the computational cost of all-atom force-fields remains too expensive for studying many interesting large-scale systems. One strategy for modeling the dynamics of folded proteins is to simplify the complicated all-atom potentials to a quadratic function in the vicinity of native state. The quadratic form allows for decomposition of the motions into vibrational modes with different frequencies, known as normal modes, and this approach has been widely used in computational studies of macromolecules since its introduction over two decades ago  \cite{go1983-3696,brooks1983-6571,levitt1985-423}. One of its advantages has been in determining the concerted motions that involve large parts of the protein, which correspond to the lowest-frequency modes. These ``global'' modes have been used to predict protein flexibility \cite{cui2004-345} and to study the mechanism of protein function where protein motion plays a key role \cite{ma1997-114}.  

Coarse-grained models, which are based on a simplified representation of protein structure, have been used historically to study the physics of folding and conformation changes in biomolecules \cite{mccammon1976-325}.  They remain attractive today, despite the exponential growth in computing power, because both the size of molecular structures being determined and the volume of structural data has increased at a similar rate.  A class of simple coarse-grained models known as Elastic Network Models, which are based on Hookean spring interactions, has been in use for a decade \cite{tirion1996-1905}, with some success at capturing features of protein dynamics  \cite{bahar2005-586}. These models define spring-like interactions between residues closer than a certain cutoff distance, which gives good agreement with overall flexibility profiles for protein structures. 

X-ray crystallography has been responsible for determination of the vast majority of protein structures to date. Conformational changes can also be observed, for instance from multiple structures of a structure under  different conditions, or as multiple conformations within a single crystal seen in high-resolution structures.  Crystallography also provides a measure of mobility through refinement of Debye-Waller temperature factors, or B-factors, for individual atoms.  This parameter is a measure of uncertainty in atomic position, and incorporates model error, lattice defects, and other experimental sources of noise in atomic position, in addition to positional variance due to internal protein motion.  The noise contributions to the B-factor are large in low resolution structures, but are far less prominent in well-refined high-resolution crystal data. Numerous studies have found good agreement between the B-factors and other experimental dynamic measures, as well as with computational predictions from molecular dynamics simulations.

The study of protein conformational dynamics benefits from an interplay between experimental data and computational modeling.  A number of studies have compared the predictions of directionality and magnitude of motion from normal mode analysis with observed conformational changes, but typically the studies have focused on individual structures.  Only recently have the computational capabilities advanced to easily process large data sets and sufficient experimental data has been amassed to perform large, systematic validations of computational models of protein dynamics.  Gerstein and co-workers \cite{alexandrov2005-633} have compared the predictions of directionality of motion with 377 structures of proteins in two conformations.  Teasdale and co-workers \cite{yuan2005-905} predicted B-factors from sequence information over a set of 766 protein chains. Halle computed residue flexibility from packing density considerations for a set of 38 structures, and compared them with B-factors \cite{halle2002-1274}.  Zhou and co-workers used an all-atom model developed for studying folding pathways to predict the flexibility of 18 structures \cite{pandey2005-1772}.  The predictions of a residue-level elastic network model called the Gaussian Network Model (GNM) \cite{bahar1997-173} were systematically tested on a set of B-factors from 113 crystal structures \cite{kundu2002-723}, and found that GNM performed substantially better than a rigid-body model of protein motion. 

This paper presents a systematic extension of GNM by incorporating chemical information into the coarse-grained model.  We optimized and validated this model, called the Chemical Network Model, using a data set of B-factors from 98 of the highest resolution crystal structures in the Protein Data Bank.  We test the effect of stepwise addition of several chemical parameters, and increase its complexity until no further gains in predictive power are obtained. 

\section*{Theory and Methods}

\begin{figure}[htbp]
\begin{center}
\vspace{.2in}
\centerline {
\includegraphics[width=6in]{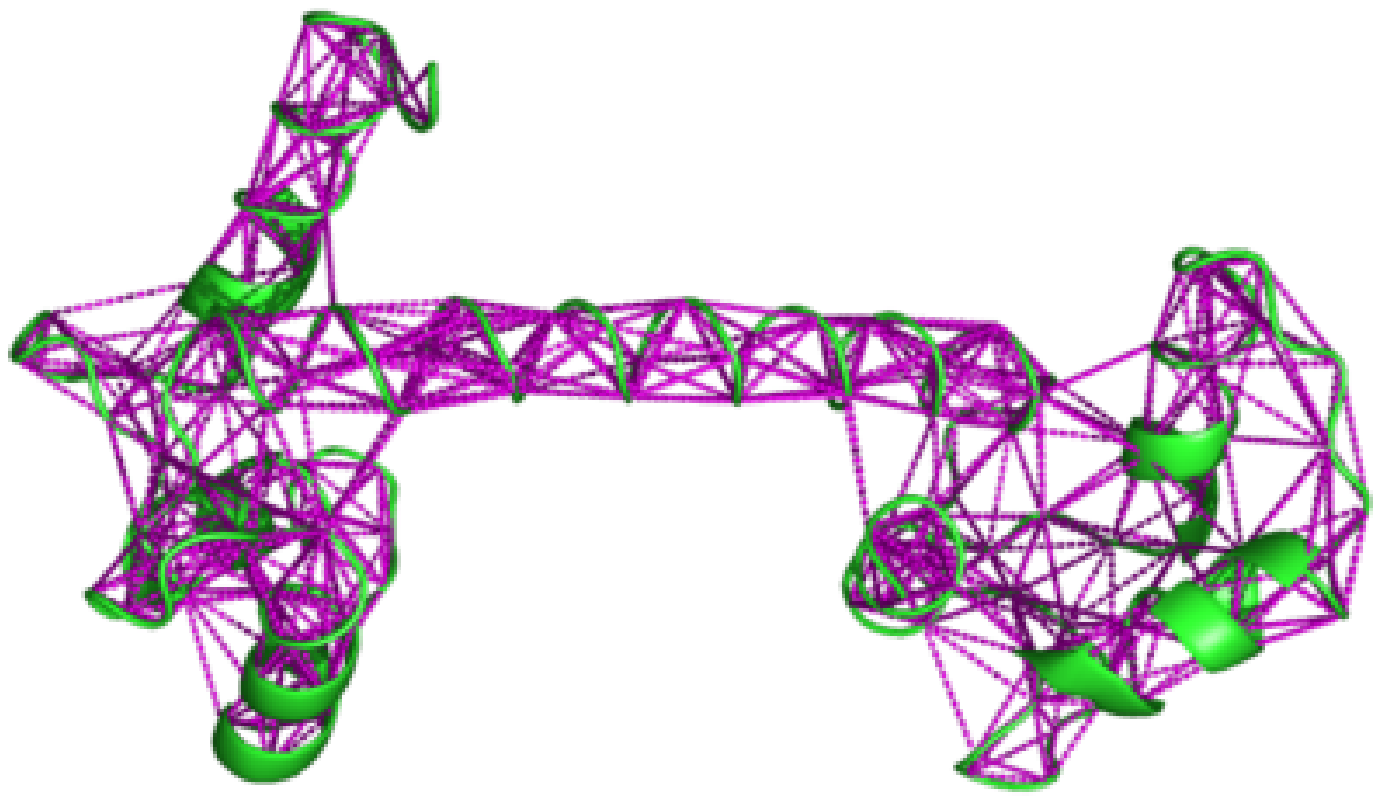}
}
\vspace{.2in}

\caption{}
\label{fig:calm_gnm}
\end{center}
\end{figure}

The Gaussian Network Model (GNM) has been described in detail elsewhere \cite{bahar1997-173}; briefly, it defines a potential based on distance between C$\alpha$ atoms.  Residue pairs within a cutoff distance $R_c$ are connected by Hookean spring potentials (Figure \ref{fig:calm_gnm}). The resulting Hessian, also known as the Kirchoff matrix, contains diagonal elements equal to the number of contacts for residue $i$, while the off-diagonal elements are -1 if there is a contact between residues $i$ and $j$. If $R_{ij}$ is the distance between C$\alpha$s of residues $i$ and $j$, then the Hessian matrix elements are defined as follows:
\begin{align}
H_{ij} & = 
\begin{cases} -1  & \text{if $R_{ij} \le R_c$} \\
 0 &  \text{if $ R_{ij} > R_c$}
\end{cases} \\
H_{ii}  & =  - \sum_j H_{ij}  \notag
\label{eq:gnm}
\end{align}

We modify and extend the Gaussian Network Model in two ways.  First, we define residue contact based on the closest distance between nonhydrogen atoms of the two residues, instead of only considering C$\alpha$ atoms.  Thus, we use the positions of all atoms to determine the interaction potential at the residue level. Second, we introduce different classes of residue interactions, with distinct Hookean spring constants. If $H_a$ is the Kirchoff (contact) matrix for class $a$, the total Hessian matrix for the harmonic model is a linear combination of the matrices, with $k_a$ as the interaction constant for each class, for example: 
\begin{equation}
H_{total} = k_{covalent}H_c + k_{polar}H_p + k_{nonpolar}H_n + k_{mixed} H_m 
\label{eq:sum_hess}
\end{equation}
The constants are determined by fitting predicted fluctuations against a data set of crystallographic B-factors, as described below.  The total Hessian is then diagonalized to find the normal modes, or eigenvectors $u_i$ and the corresponding frequencies $\omega_i$: $Hu_i = \omega_i^2 u_i$. The decomposition allows us to compute both self- and cross-correlation of motion between residues from the covariance matrix, which is proportional to the pseudo-inverse of the Hessian \cite{go1990-105}. Specifically, we are interested in the positional variances, or the mean square fluctuations of residues, which are determined as follows ($\Delta x_{i}$ is the deviation of position of residue $i$ from the mean and $u_{ij}$ is the $j$-th element of the $i$-th normal mode):

\begin{equation}
<\Delta x_{i}^2> = \sum_j \frac{1}{\omega_i^2} u^2_{ij} 
\label{eq:msf}
\end{equation}
Note that the modes with the lowest frequencies make the greatest contribution to residue mobility, so a small fraction of all the modes is sufficient to obtain a good approximation of the sum.

We used perl programs to parse PDB files and determine residue contact matrices based on atomic coordinates.  To determine the optimal cutoff parameters, a range of C$\alpha$ cutoff distances was used, from 6 to 12 \AA, similarly, nearest-atom cutoff distances were varied from 3.5 to 9 \AA.  Copies of the protein molecule surrounding the structure in the crystal were generated using the symexp command in PyMOL \cite{pymol-0}.  In both GNM and CNM, the crystal environment was taken into account by adding interactions between residues involved in crystal contacts, without explicitly adding crystal copies to the matrix.  Since the model does not contain directional information, this is a more precise approximation of the effect of the crystal lattice than explicit inclusion of the first layer of crystal neighbors.  For the nearest-atom method, the interaction of atoms in more than position were counted proportional to their occupancy. The matrices were diagonalized using the MATLAB computing environment \cite{matlab-0}.  The predicted mean square fluctuations (MSF) were computed as a sum over all the normal modes as shown in equation \ref{eq:msf}, and then compared to a set of experimental B-factors.

The data set was obtained by searching the Protein Data Bank for protein structures determined by X-ray crystallography to at least 1.0 \AA\ resolution, containing at least 50 residues in a single chain.  Structures with more than 50 \% identity were discarded, leaving 98 non-redundant proteins.  These are structurally diverse, representing all major SCOP families, as shown in the comprehensive table in Supplemental Materials.  Isotropic C$\alpha$ B-factors from each structure were normalized to mean 1, to enable simultaneous fitting over multiple structures.  The B-factors from atoms with more than one conformer or occupancy less than 1 were not used for fitting or validation, due to the linkage between occupancy and the B-factor. The usable data set consists of 20942 B-factors.  In addition, non-protein residues were considered for a subset of structures with ligands or cofactors other than those from crystallization buffers or precipitants.  For the 68 structures with ligands or cofactors, separate calculations were performed with and without including the non-protein molecules in the model. Each molecule, whether large or a single metal ion, was considered as a single residue and included in the Hessian, but only the B-factors from protein residues were compared with the predictions.

We determine interaction constants that maximize the correlation between computed fluctuations and crystallographic B-factors.  Since there is no analytic expression for the fluctuations as a function of the spring constants, standard gradient-based optimization techniques are not applicable, and we use parameter-space search methods. The first model consists of two classes of residue interactions: bonded and nonbonded. Because we test for correlation, scaling is immaterial, so the bonded parameter was set to 1, and only the nonbonded interaction constant was varied. We use a simple search over a range of values from 0.01 to 1 for the nonbonded constant.

We expand the model to include further chemical categories, specifically polar interactions, nonpolar interactions, and those that do not fall in either category. These were defined by the types of the nearest atoms for a residue pair.  Nitrogens or oxygens less than 3.3 \AA\ apart were classified as a polar contact.  The cutoff distance for the other two categories were varied from 3.5 \AA\ to 9 \AA: the nonpolar category, which is defined as two carbon atoms, with the exception of backbone carbons and certain charged carbons, such as those in carboxyl groups, and the mixed category, which included any other atom pairs.  To find the maximum correlation by varying the three parameters we utilized a standard parameter space method, called the simplex algorithm \cite{nr1992-0}. It involves evolving a polygonal region (simplex) in parameter space in an effort to enclose the optimal point.  The algorithm was implemented in MATLAB and applied to three training sets of 15 structures, while the remaining 53 structures served as a test set for unbiased assessment of the optimized parameters. 

\section*{Results}
\begin{figure}[htbp]
\begin{center}
\vspace{.2in}
\centerline {
\includegraphics[width=3in]{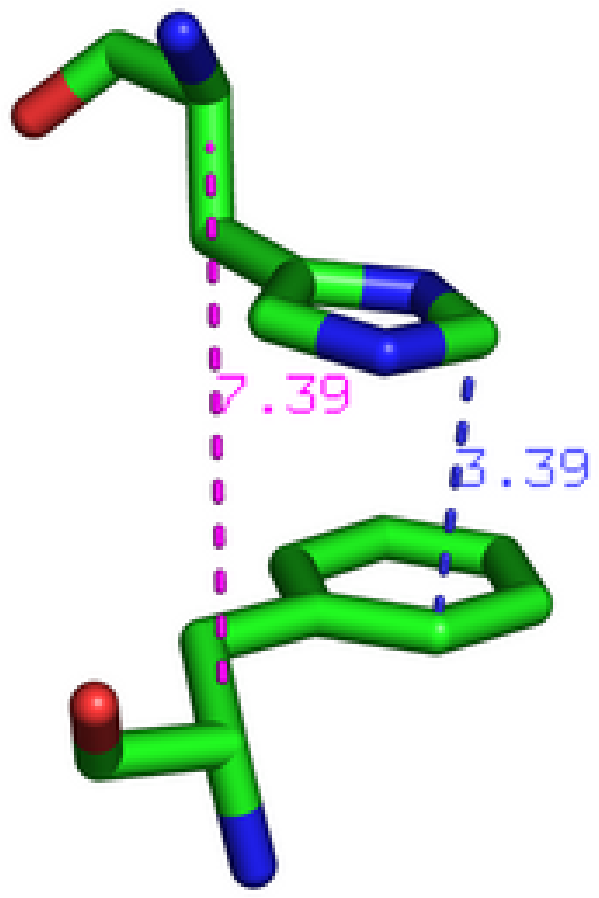}
\includegraphics[width=3in]{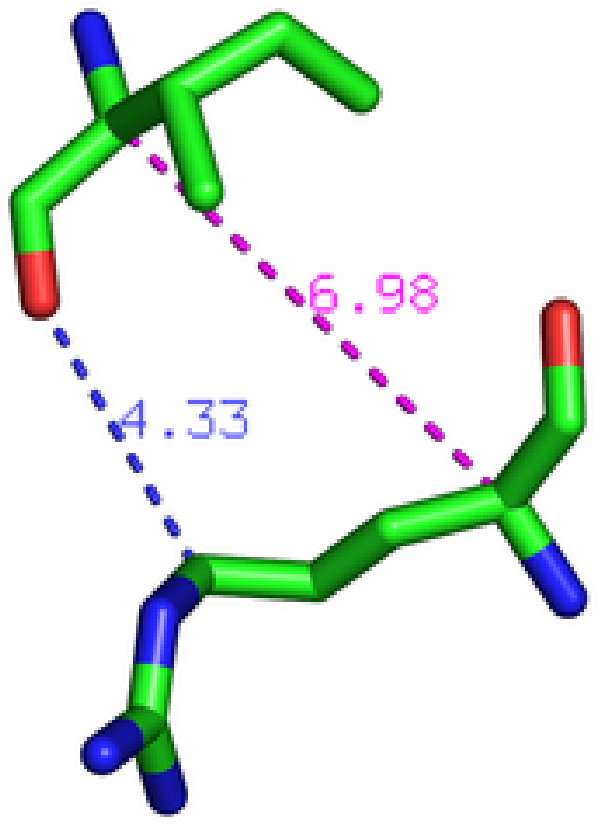}
}
\vspace{.2in}

\caption{}
\label{fig:ca_vs_atom}
\end{center}
\end{figure}

The Gaussian Network Model represents all residue interactions within a cutoff distance between C$\alpha$ atoms as identical harmonic potentials.  We introduce two modifications to the model to better represent the chemistry of residue interactions. First, interaction types are separated into classes with different strengths, or Òspring constantsÓ, to model the diversity of residue interactions in protein structures.  Second, inter-residue contacts are defined by the closest distance between atom pairs, rather than the distance between C$\alpha$ atoms. Figure \ref{fig:ca_vs_atom} demonstrates how a C$\alpha$ distance cutoff of 7.3 \AA\ can miss a strong ring-stacking interaction, but may include a weak contact instead.  While all atoms are considered in determining residue interactions, the size of the Hessian matrix produced by the model is equal to the number of residues in the structure, as in GNM.  

The results demonstrate that a combination of the two modifications results in significantly larger improvement than either one alone.  Tables \ref{tab:gnm} and \ref{tab:cnm} show the results for C$\alpha$ distance cutoff and the nearest-atom distance cutoff, respectively.   Average correlations over the entire data set were computed for a range of cutoff distances and a number of nonbonded interaction constants.  C$\alpha$ distance method benefits from separation of interaction types, especially for the larger cutoff distances, in which large numbers of contacts are included. The improvement is greatest for the combination of nearest-atom cutoff of 4 \AA\ and the nonbonded parameter of 0.1, giving an average correlation of 0.743.  This is significantly higher than the best GNM prediction of 0.643, at C$\alpha$ cutoff of 7.5 \AA. The improvement is seen in almost every structure, listed in Supplemental Materials. Thus, the combination of the two modifications, termed the Chemical Network Model (CNM), improves the prediction power by 10\%.  The GNM results are consistent with an earlier large-scale study \cite{kundu2002-723}, which found an average correlation of 0.66 at cutoff of 7.3 \AA.  As in that work, crystal contacts were included in the models, as described in Methods, and resulted in significantly improved agreement (data not shown). 

\begin{figure}[htbp]
\begin{center}
\vspace{.2in}
\centerline {
\includegraphics[width=3.5in]{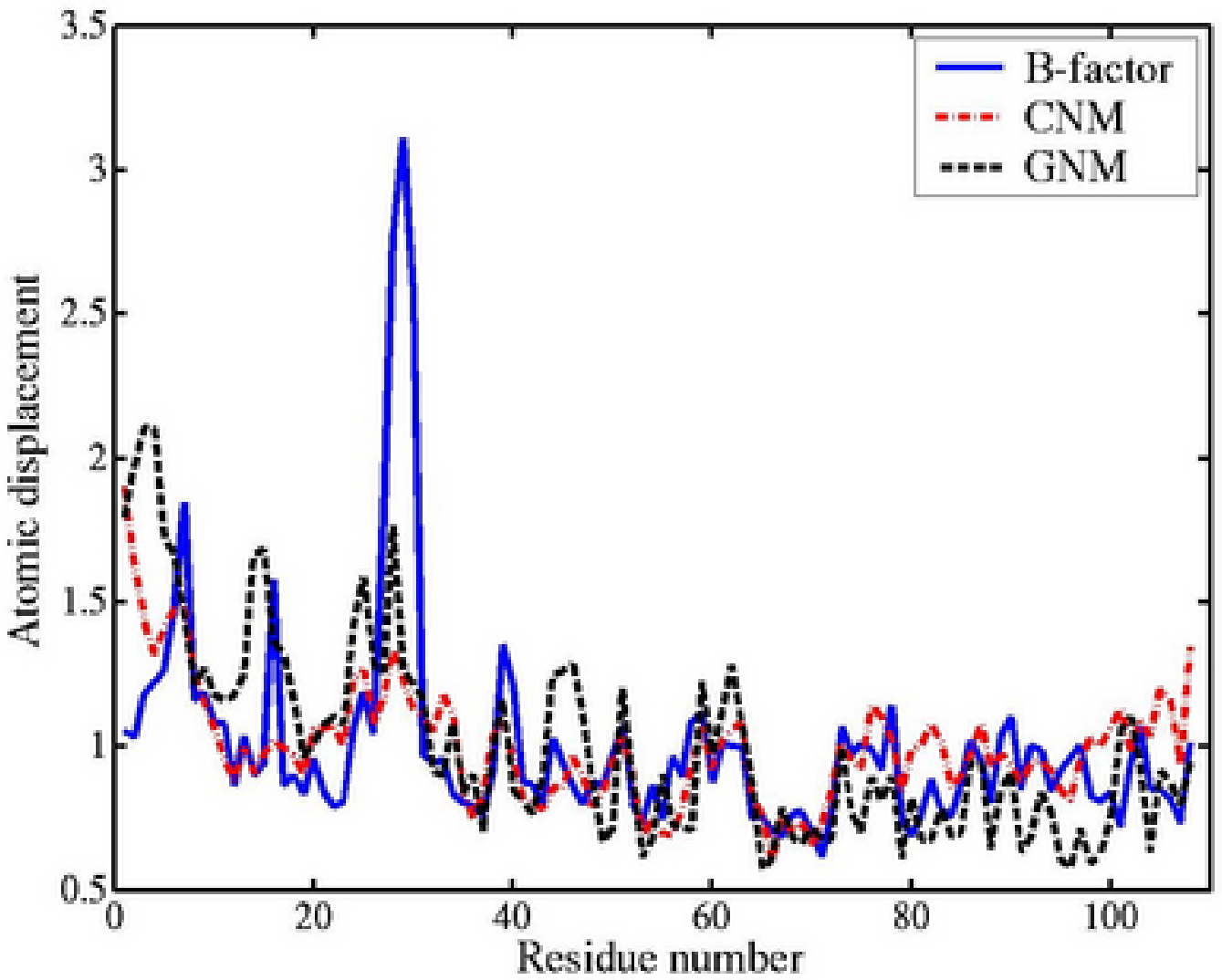}
\includegraphics[width=3.5in]{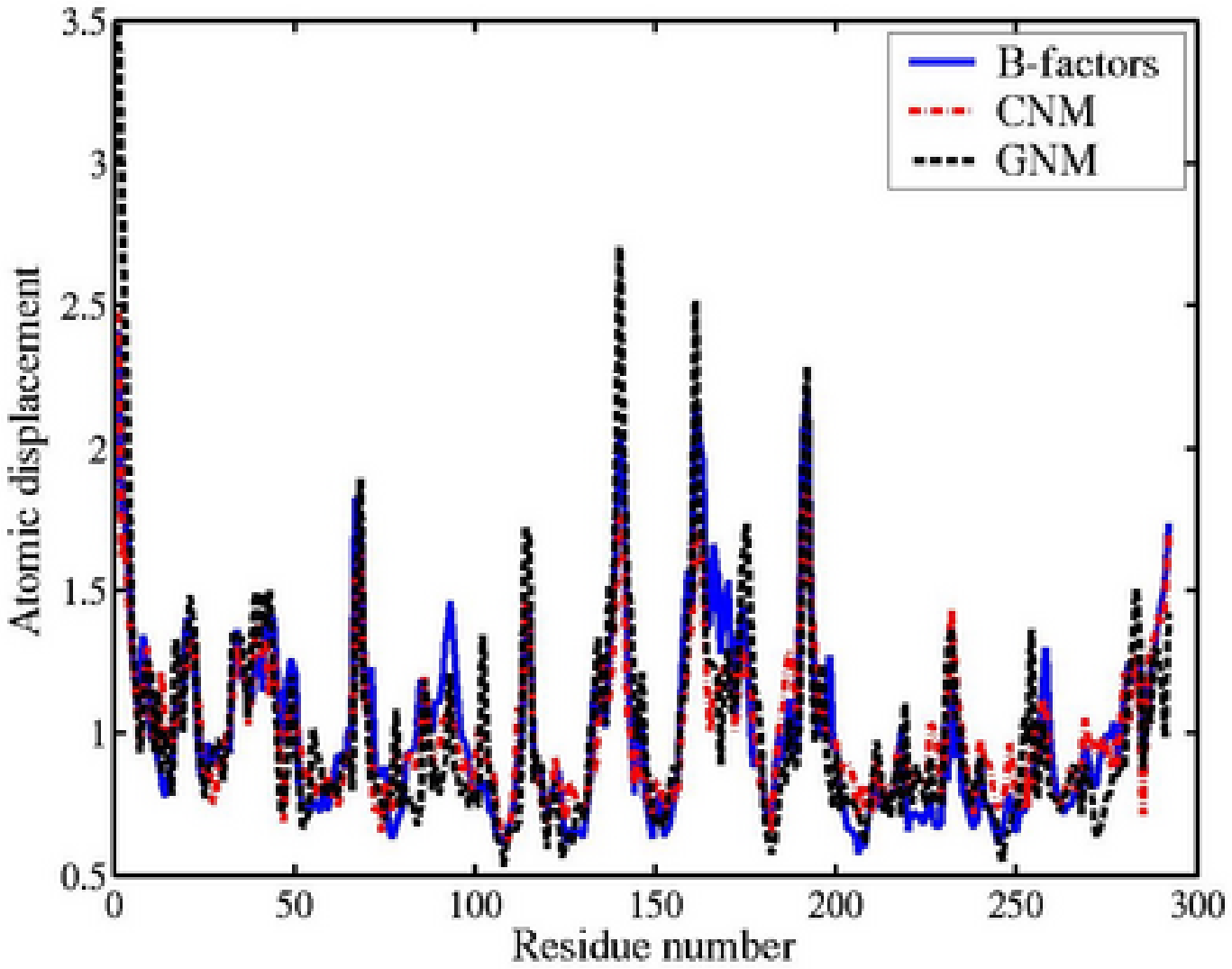}
}
\vspace{.2in}

\caption{}
\label{fig:best_worst}
\end{center}
\end{figure}

In both GNM and CNM results, there is considerable variation in fluctuation prediction over different structures.  
One hypothesis is that the elastic network models are best suited for dense, globular structures, and are less accurate for sparsely packed residues on protein surface \cite{vanwynsberghe2005-2939}.  Table \ref{tab:surface} presents a breakdown of results for structures with different fraction of surface residues, defined as those with less than 3 nonbonded contacts in CNM. We see that structures with the lowest and highest fraction of surface residues show significantly lower average correlations in GNM and CNM. Contrary to expectation, the structures with the lowest fraction of surface residues have the worst predictions in both models, but also show the greatest improvement from GNM to CNM (from 0.49 to 0.64). The average correlation is also significantly lower in the set with the highest surface fraction, and the standard deviation of prediction quality is also higher in the high and low surface fraction sets. Figure \ref{fig:best_worst} illustrates the variability of model agreement with plots of normalized fluctuation profiles and the B-factors for the two structures with the best and the worst correlation with CNM predictions.  PDB structure 1J0P, with CNM correlation of 0.46, is a small bacterial cytochrome C3 with 4 embedded hemes, and due to this has the inordinately high fraction of surface residues of 0.31.  On the other hand, the best prediction is seen in PDB ID 2BW4, with CNM correlation at 0.9  This is a nitrite reductase that has a well-packed globular fold, with the exception of a long C-terminal tail that is packed by crystal contacts, with overall surface fraction of 0.10.  In addition, we observe a positive effect of larger protein size on prediction quality for both methods, as shown in table \ref{tab:size_matters}.  

We extend the classification of residue interactions by separating the nonbonded category into polar, nonpolar, and mixed.  Separate Kirchoff matrices were computed for each category, and optimal interaction constants for each were found by the simplex method, as described in Methods.  3 training sets of 15 structures were used for optimization, and fluctuation predictions were computing using the optimal parameter sets, and compared on a separate test set of 53 residues; the results are shown in Table \ref{tab:optim_results}.  Although some improvement can be seen from optimization on the test sets, and larger improvement can be seen in individual structures (data not shown), the correlation over the reference set using the optimized parameters is lower than that with all the nonbonded parameters set to 0.1.  The increase in correlation in individual sets is only a result of fitting imperfections of the coarse-grained model for particular structures, not evidence of general differences in interaction strength.  Thus, there is not a sufficient distinction in the different types of interactions to warrant including them as separate categories.

Several other factors were considered in order to further improve the prediction quality. The presence of cofactors or ligands in the crystal structures can affect the mobility of the neighboring residues.  The results presented above omitted the non-protein residues, and when the subset of structures with ligands or cofactors is compared to those without, the group without non-protein residues has a slightly higher average correlation.  Addition of cofactors and ligands, as described in Methods, improves the average correlation for the ligand-containing group from 0.740 to 0.748, similar to the value of 0.749 for the ligand-free group. Thus, the consideration of non-protein residues results in a small but measurable improvement in mobility prediction. It also behooves us briefly to report the modifications of the model that either yielded no improvement or were detrimental to the prediction quality. They include making the interaction between residues proportional to the number of atom pairs within interaction range, adding mass-weighting to the Hessian matrix, and introducing a new interaction category for residues within the same secondary structure element.

\begin{figure}[htbp]
\begin{center}
\vspace{.2in}
\centerline {
\includegraphics[width=5in]{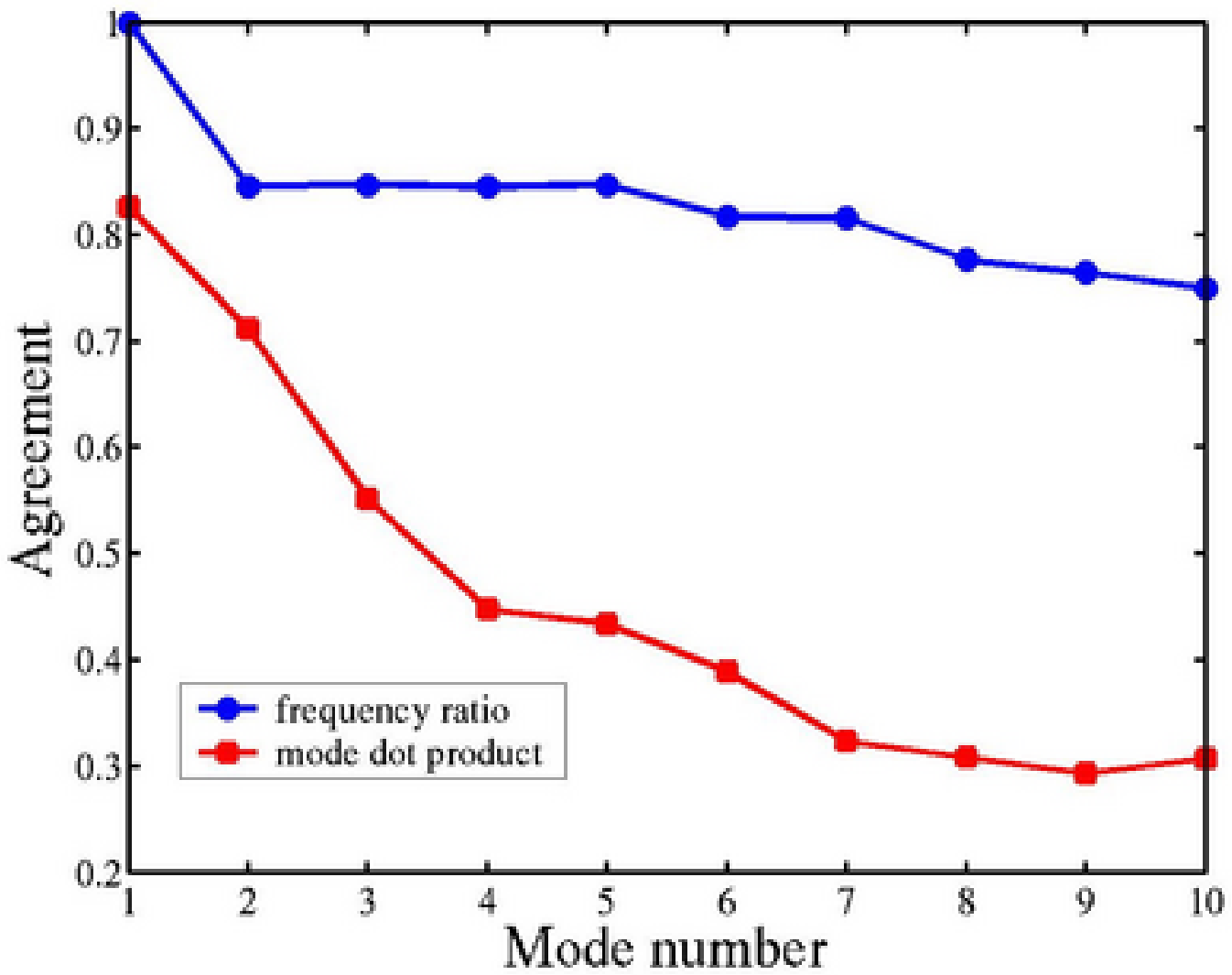}
}
\vspace{.2in}

\caption{}
\label{fig:modes_freq}
\end{center}
\end{figure}

The lowest-frequency normal modes and their eigenvalues from GNM and CNM were compared.  Figure \ref{fig:modes_freq} shows the mean dot product between corresponding normal modes and the ratio of the eigenvalues, normalized to the lowest value.   We see that the lowest-frequency modes are quite similar, but progressively diverge at higher frequencies, with little similarity remaining by normal mode 10.  This demonstrates that the two methods share an overall gross structure, which is reflected in the lowest-frequency modes, but the details of contact selection and interaction strengths play a greater role at higher frequencies. Still, the differences are not negligible, and the improved predictive power of CNM suggests that its normal modes are more accurate, as well. 

\section*{Discussion}
Simple models of complex systems serve at least two purposes. Practically, they offer efficient computation, enabling approximate treatment of objects that are beyond the current computational capabilities of more realistic methods. For instance, the dynamics of large macromolecular assemblies are still prohibitively expensive to be treated by all-atom molecular dynamics. Coarse-grained potentials provide an opportunity to quantitatively study systems such as viral capsids \cite{tama2005-299} and the ribosome \cite{wang2004-302}, which play critical biological roles.  The second advantage of simple modeling is that it sharpens our understanding.  Beginning with the most basic assumptions and gradually adding details, one can arrive at a minimal set of key variables that describe an opaque reality. This was the approach taken by this study.

The Gaussian Network Model has been successful at predicting the features of collective protein motions, as evidenced by comparison of fluctuation profiles with crystal B-factors and NMR relaxation data \cite{haliloglu1999-654}, as well as by prediction of conformational changes from low-frequency normal modes \cite{temiz2004-468}.  A previous large-scale study \cite{kundu2002-723} has systematically assessed its agreement with crystal B-factors, finding an average correlation of 0.66, while a rigid-body model obtained a correlation of 0.52.  This provided clear evidence that the contact topology of protein structures plays a key role in determining the near-native dynamics. However, there is room for improvement of the correlation coefficient, and this motivated our chemistry-based coarse-grained model of protein dynamics. 

The Chemical Network Model rests on the assumption that atomic contacts are the primary means of inter-residue interaction. We construct the Hessian matrix at the residue level from atomic information present in crystal structures. Further, we divide the interactions into classes, first into bonded and nonbonded, and then split the nonbonded category.  Simplified residue-level forcefields which distinguish different interaction types have been proposed before, ranging from Go-like models for studying folding pathways \cite{portman1998-5237, micheletti2001-088102} to amino-acid specific potential of Miyazawa and Jernigan \cite{miyazawa1996-623}.  In contrast, our model applies to vibrational fluctuations in the native state, and is distinguished from these models by its simplicity and the systematic comparison against a large data set of reliable measurements of protein mobility. Similar modifications of elastic network models were reported very recently: one that strengthened the bonded interactions in GNM to match the predictions of all-atom normal mode analysis \cite{ming2005-198103}, and another \cite{jeong2006-296} which divided interactions into several types ranging from disulfide bonds to van der Waals contacts to construct an extension of the anisotropic version of GNM, known as ANM \cite{atilgan2001-505}.  However, the first study uses a C$\alpha$-cutoff potential, and we demonstrated that the combination of nearest-atom contact potential and different interaction strengths leads to further improvement.  The second study did not justify the values of parameters chosen for the different interaction types. Finally, both use only a few examples rather than a large data set to validate their models.

Our results show that the nearest-atom contact potential coupled with differentiation of bonded and nonbonded interactions leads to a synergistic improvement of mobility prediction.  The nearest-atom contact potential adds some contacts missed by GNM, yet excludes other GNM interactions. On average, there are fewer residue contacts in CNM with the nearest-atom cutoff of 4 \AA\  than in GNM with the optimal cutoff of 7.5 \AA. The improvement of contact selection is apparently counterbalanced by a reduction in contact density, which may be why nearest-atom contact potential alone has no significant effect on prediction quality.  The introduction of bonded and nonbonded constants modifies the relative density of contacts to better match the observed residue mobility.  We also observe that both GNM and CNM work best for typical globular structures, and those with very high or very low fraction of surface residues show substantially lower prediction quality.  This may also explain why larger proteins tend to show better prediction, since the surface fraction is more stable, and illustrates the suitability of coarse-grained modeling for large macromolecular assemblies.

Classifying the nonbonded interactions into polar, nonpolar, and mixed, did not yield improvement in an unbiased comparison with a reference set of 53 structures.  The correlation coefficient is relatively insensitive to changes in the interaction parameters: an order of magnitude change between bonded and nonbonded parameters was required to achieve a 10\% improvement in average correlation, and smaller tune-ups of the nonbonded parameters have no significant effect.  Although optimization produces substantial improvement in individual structures (data not presented), these optimizations are apparently not applicable across a wide array of structures.  

The failure of the more complex model illustrates both the strengths and the limitations of the coarse-grained elastic network model.  Addition of simple chemical information, together with consideration of crystal contacts and co-crystallized ligands and cofactors produces the average correlation of 75\% with experimental data, with even better agreement for larger structures. This is solid quantitative predictive power for a model at the residue level, and better agreement probably requires detailed atomic modeling.  The coarseness of the model also leads to its limitation: addition of more information is washed out due to the scale. This suggests that this class of models is unsuitable for addressing some important questions, such as the effect of mutations on protein motion \cite{bae2006-2132}, which sometimes have a direct functional link \cite{wong2005-6807}.

Prediction of observed fluctuations is only a means of validating the model, not a goal in itself.  While computation of average positional deviation is sometimes useful, the most promising applications of harmonic models have been the use of low-frequency modes to study persistent collective motions in protein structures.  This information has been used for prediction of mechanisms of functionally significant motions \cite{ma1997-11905,cui2004-345, vanwynsberghe2004-13083} or in quantifying allosteric interaction between distant parts of a protein structure \cite{ming2005-697}.  Normal modes have enable the improvement of crystallographic structure determination by molecular refinement \cite{suhre2004-796}, the refinement of low-resolution structures of large assemblies \cite{tama2004-985,mitra2005-318}.
Coarse-grained normal modes are also useful in analyzing the large numbers of newly determined structures, for instance in the prediction of active sites \cite{yang2005-893}, automated decomposition of protein structures into domains \cite{kundu2004-725}, and a determination of networks of residues involved in key conformational changes \cite{zheng2005-565}.  While CNM and GNM predict similar lowest-frequency modes, the improvement in fluctuation prediction suggests that the changes in the modes are significant, and may provide more accurate prediction of collective motions, especially for large protein structures and assemblies.

\section*{Conclusion}
We have extended GNM by constructing the Hessian contact matrix based on atomic contacts, and separating residue interactions into bonded and nonbonded. The resulting Chemical Network Model shows considerable improvement of the prediction of crystallographic B-factors, giving 75\% average correlation on a data set of 98 ultra-high resolution structures. However, further separation of nonbonded interactions into polar, nonpolar, and mixed, did not yield any improvement in correlation coefficient. We have improved the residue-level elastic network model without increasing the computational cost, and found an appropriate level of complexity for the application.

\begin{acknowledgments}
D.A.K. was supported in part through an National Library of Medicine training grant to the Computation and Informatics in Biology and Medicine program at UW-Madison (NLM 5T15LM007359). Q.C. is an Alfred P. Sloan Research Fellow.
\end{acknowledgments}

\section*{References}
\bibliography{bibkon}

\newpage
\section{Tables}
\begin{table}[htbp]
   \centering
   \topcaption{Average correlation of B-factor prediction for C$\alpha$ distance cutoff models. The cutoff distance is varied across the columns, and the nonbonded interaction parameter varies by row; the highest correlation for each cutoff value is in bold} 
   \begin{tabular}{@{} l c c c c c c c c c@{}} 
      \toprule
  	non		 & 6 \AA 		& 6.5 \AA		& 7 \AA 		& 7.5 \AA		& 8 \AA 		& 9 \AA 		& 10 \AA 		& 11 \AA 		& 12 \AA \\ 
        \hline
         	  1.0    & 0.542 		 & 0.578 		& 0.624 		& 0.643 		& 0.629 		& 0.619 		& 0.627 		& 0.634 		& 0.628 \\
		  0.5    &\textbf{0.552}& 0.604		& 0.639 		& 0.655 		& 0.643 		& 0.630 		& 0.636 		& 0.641 		& 0.633 \\ 
		  0.25  & 0.548 		&\textbf{0.615} & \textbf{0.646}&\textbf{0.662} & 0.655 		& 0.645 		& 0.649 		& 0.652		& 0.643 \\
 		  0.15  & 0.540 		& 0.610 		& 0.643 		& 0.661		& \textbf{0.659} & 0.656 		& 0.660 		& 0.662 		& 0.652 \\   
 		  0.1    & 0.525 		& 0.597 		& 0.634 		& 0.654 		& 0.658 		& 0.661 		& 0.668 		& 0.670 		& 0.660 \\ 
 		  0.05  & 0.490 	        & 0.562 		& 0.605 		& 0.631 		& 0.643 		& \textbf{0.662} & \textbf{0.676} & \textbf{0.682} & \textbf{0.672} \\   
 		  0.01  & 0.395		& 0.451		& 0.500		& 0.533		& 0.558 		& 0.608 		& 0.646 		& 0.668 		& 0.670 \\   
     \toprule
   \end{tabular}
   \label{tab:gnm}
\end{table}

\newpage
\begin{table}[htbp]
   \centering
  \topcaption{Average correlation of B-factor prediction for nearest-atom distance cutoff models. The cutoff distance is varied across the columns, and the nonbonded interaction parameter varies by row; the highest correlation for each cutoff value is in bold.} 
   \begin{tabular}{@{} l c c c c c c c c@{}} 
      \toprule
         non 		& 3.5 \AA		& 4 \AA		& 4.5 \AA		& 5 \AA 		& 6 \AA 		& 7 \AA 		&  8 \AA 		& 9 \AA \\ 
        \hline
         	1.0    & 0.569 		& 0.649 		& 0.644 		& 0.632 		& 0.630 		& 0.625 		& 0.639 		& 0.633\\
		0.5    & 0.612 		& 0.685 		& 0.676 		& 0.662 		& 0.652 		& 0.637 		& 0.648 		& 0.640\\ 
		0.25  & 0.642 		& 0.717 		& 0.707 		& 0.692 		& 0.677 		& 0.656		& 0.661 		& 0.651\\
 		0.15  & \textbf{0.649}& 0.735 		& 0.726 		& 0.713 		& 0.696 		& 0.673 		& 0.674 		& 0.662\\   
 		0.1    & 0.642		& \textbf{0.743} & 0.737 		& 0.725		& 0.709 		& 0.688 		& 0.686 		& 0.672\\ 
 		0.05  & 0.611 		& 0.735 		& \textbf{0.738}& \textbf{0.731}& \textbf{0.721}& \textbf{0.709}& \textbf{0.706}& 0.691\\   
 		0.01  & 0.497		& 0.625		& 0.650		 & 0.654 		& 0.669 		& 0.692 		& 0.704		& \textbf{0.697}\\   
     \toprule
   \end{tabular}
   \label{tab:cnm}
\end{table}

\newpage
\begin{table}[htbp]
  \centering
   \topcaption{Fraction of surface residues and accuracy of prediction} 
   \begin{tabular}{@{} l c c c c @{}} 
   \toprule
	Surface: & Low & Medium & High & Total \\ 
       \hline
       	structures & 10 & 78 & 10 & 98 \\
	residues & 1093 & 18120 & 1730 & 20942 \\ 
	surface fraction & 0.049 & 0.103 & 0.184 & 0.107 \\
 	GNM \footnotemark[1] & 0.495$\pm$0.107 & 0.657$\pm$0.095 & 0.592$\pm$0.099 & 0.643$\pm$0.105 \\   
 	CNM \footnotemark[1]  & 0.648$\pm$0.111 & 0.752$\pm$0.082 & 0.709$\pm$0.099 & 0.743$\pm$0.089\\ 
      \toprule
         \footnotetext[1]{average and standard deviation of correlation over the set}
    \end{tabular}
   \label{tab:surface}
\end{table}

\newpage
\begin{table}[htbp]
   \centering
   \topcaption{Effect of protein size on average correlation with GNM and CNM fluctuations} 
   \begin{tabular}{@{} l r r r r r @{}} 
   \toprule
	$>$size & 500 & 300 & 200 & 100 & all \\ 
       \hline	
	structures & 4  & 20 & 48 & 73 & 98 \\ 
 	residues & 2828  & 8804 & 15731 & 19277 & 20942 \\   
 	GNM & 0.641 & 0.660 & 0.664 & 0.651 & 0.643 \\ 
 	CNM & 0.774 & 0.765 & 0.760 & 0.746 & 0.743 \\   
      \toprule
    \end{tabular}
   \label{tab:size_matters}
\end{table}


\newpage
\begin{table}[htbp]
   \centering
   \topcaption{Optimization of nonbonded interaction parameters over 3 training sets of 15 structures and cross-validation on a reference set of 53 structures.} 
   \begin{tabular}{@{} l c c c @{}} 
   \toprule
	training set & Set 1 & Set 2 & Set 3  \\ 
       \hline
       residues & 2249 & 3805 & 3229 \\	
	polar\footnotemark[1]       & 0.115& 0.147 & 0.129 \\ 
 	nonpolar\footnotemark[1] & 0.106 & 0.107 & 0.049 \\   
 	mixed\footnotemark[1]      & 0.123 & 0.072 & 0.045  \\ 
 	training before\footnotemark[2]  & 0.701 & 0.740 & 0.705 \\   
 	training after\footnotemark[2]   & 0.702 & 0.752 & 0.726 \\ 
	reference before\footnotemark[2]  &  0.761 & 0.761 & 0.761  \\
	reference after\footnotemark[2]    &  0.757 & 0.754 & 0.740 \\  
      \toprule
   \footnotetext[1]{optimal parameter values for the training set as found by the simplex method}
   \footnotetext[2] {average correlations with all parameters at 0.1 (before) and with optimized parameters (after)}
    \end{tabular}
   \label{tab:optim_results}
\end{table}

\newpage
\section{Figure Captions}
Figure \ref{fig:calm_gnm}: Cartoon of calmodulin structure (1EXR) in green with C$\alpha$ atoms within 7.3 \AA\ connected by magenta dotted lines to represent GNM interactions. \\ \\

Figure \ref{fig:ca_vs_atom}: Contrast between residue interactions selected by C$\alpha$ distance (magenta) and nearest-atom distance (blue). A: residues with a strong ring-stacking interaction with C$\alpha$ distance greater than 7.3 \AA. B: residues not in chemical contact with C$\alpha$ distance less than 7 \AA. Both examples from sperm whale myoglobin structure (1A6M). \\ \\

Figure \ref{fig:best_worst}: Examples of computed fluctuation profiles and experimental B-factors (normalized) A:  Worst prediction, 1J0P (0.46 CNM, 0.46 GNM) B: Best prediction, 2BW4 (0.9 CNM, GNM 0.84). \\ \\ 

Figure \ref{fig:modes_freq}: Comparison of corresponding low-frequency modes from GNM and CNM. The blue curve shows the ratio (lower to higher) of the frequencies normalized to the lowest frequency, averaged over the 98 structures. The red curve is the average dot product between the corresponding normal modes. Note the fast decline of the normal modes at higher frequencies.

\newpage
\section{Figures}

\newpage

\newpage

\newpage

\end{document}



\title{Supplemental Materials for: Optimization and evaluation of a coarse-grained model of protein motion using X-ray crystal data}

\author{Dmitry A. Kondrashov$^*$, Qiang Cui$^\dagger$, and George N. Phillips, Jr.$^{\dagger,\ddagger}$}
\affiliation{%
$^*$Department of Biochemistry, \\ $^\dagger$ Department of Chemistry and Theoretical Chemistry Institute, \\
$\ddagger$ Departments of Biochemistry and Computer Science \\
University of Wisconsin - Madison, Madison, WI 53706
}%
 \email{phillips@biochem.wisc.edu}

\date{\today}
 
 \maketitle
 \section{Table of all structures}
   \begin{longtable}{ l c c c c c c} 
   \toprule
	PDB ID & SCOP family &  all residues & usable residues & surface fraction & GNM (7.5 \AA) & CNM (4.0 \AA) \\ 
       \hline
1a6m&a.1.1.2   & 151& 151&0.079&0.482&0.643\\
1aho&g.3.7.1   &  64&  62&0.145&0.719&0.749\\
1brf&g.41.5.1  &  53&  53&0.151&0.472&0.604\\
1byi&c.37.1.10 & 224& 210&0.105&0.611&0.724\\
1c75&a.3.1.1   &  71&  69&0.072&0.566&0.649\\
1c7k&d.92.1.1  & 132& 132&0.068&0.508& 0.73\\
1cex&c.69.1.30 & 197& 197&0.102& 0.75&0.766\\
1dy5&d.5.1.1   & 248& 248&0.105&0.673&0.639\\
1ea7&c.41.1.1  & 310& 305&0.089&0.662&0.683\\
1eb6&d.92.1.12 & 177& 177&0.051&0.572&0.799\\
1exr&a.39.1.5  & 146& 109&0.165&0.717& 0.68\\
1f94&g.7.1.1   &  63&  63&0.079&0.486&0.619\\
1f9y&d.58.30.1 & 158& 156&0.154&0.441&0.556\\
1g4i&a.133.1.2 & 123& 112&0.107&0.693&0.769\\
1g66&c.69.1.30 & 207& 205&0.088&0.714&0.797\\
1g6x&g.8.1.1   &  58&  55&0.164&0.777&0.886\\
1ga6&c.41.1.2  & 371& 369&0.133&0.738& 0.86\\
1gci&c.41.1.1  & 269& 260&0.085&0.724&0.792\\
1gkm&a.127.1.2 & 509& 508&  0.1&0.495&0.674\\
1gqv&d.5.1.1   & 135& 135&0.104&0.599&0.703\\
1gvk&b.47.1.2  & 243& 222&0.104& 0.79&0.893\\
1gwe&e.5.1.1   & 498& 491&0.075&0.657&0.757\\
1hj9&b.47.1.2  & 223& 215&0.163&0.503&0.751\\
1i1w&c.1.8.3   & 302& 299&0.057&0.479&0.663\\
1ic6&c.41.1.1  & 279& 279&0.097&0.709&0.716\\
1iqz&d.58.1.4  &  81&  81&0.148&0.648&0.809\\
1iua&g.35.1.1  &  83&  81&0.074&0.406& 0.63\\
1ix9&a.2.11.1/d.44.1.1 & 410& 381&0.097&0.585& 0.72\\
1ixh&c.94.1.1  & 321& 321&0.112& 0.68& 0.74\\
1j0p&a.138.1.1 & 108& 108&0.306& 0.48&0.464\\
1jfb&a.104.1.1 & 399& 375& 0.12&0.632&0.681\\
1k4i&d.115.1.2 & 216& 216&0.102&0.574&0.603\\
1k5c&b.80.1.3  & 333& 330&0.082&0.693& 0.78\\
1kth&g.8.1.1   &  58&  58&0.172&0.658&0.899\\
1kwf&a.102.1.2 & 363& 344&0.081&0.639&0.672\\
1l9l&a.64.1.1  &  74&  74&0.041&0.324&0.518\\
1lkk&d.93.1.1  & 109& 109&0.101&0.535& 0.64\\
1lni&d.1.1.2   & 192& 165&0.164&0.568&0.659\\
1lug&b.74.1.1  & 259& 258&0.097&0.644&  0.7\\
1m1q&a.138.1.3 &  90&  74&0.311&0.276& 0.56\\
1m40&e.3.1.1   & 263& 118&0.085&0.713&0.824\\
1mc2&a.133.1.2 & 122& 122&0.066& 0.58&0.514\\
1mj5&c.69.1.8  & 297& 266&0.079&0.668&0.757\\
1mn8&a.61.1.6  & 379& 379& 0.09&0.652&0.806\\
1muw&c.1.15.3  & 386& 341&0.073&0.701&0.724\\
1mwq&d.58.4.7  & 194& 194&0.103&0.414&0.501\\
1n4w&c.3.1.2/d.16.1.1 & 498& 452&0.104& 0.69&0.816\\
1n55&c.1.1.1   & 249& 244& 0.07&0.805&0.814\\
1nki&d.32.1.2  & 268& 262&0.141&0.712&0.744\\
1nls&b.29.1.1  & 237& 236&0.136&0.472& 0.86\\
1nqj&b.23.2.1  & 210& 210&0.114&0.766&0.744\\
1nwz&d.110.3.1 & 125& 109&0.064&0.401&0.507\\
1o7j&c.88.1.1  &1300&1296&0.097&0.596&0.749\\
1oai&a.5.2.3   &  68&  68&0.015&0.628&0.808\\
1od3&b.18.1.10 & 131& 131&0.099&0.547&0.774\\
1oew&b.50.1.2  & 329& 324&0.136&0.655&0.754\\
1ok0&b.5.1.1   &  74&  65&0.062&0.732&0.753\\
1p1x&c.1.10.1  & 501& 501&0.096&0.771& 0.87\\
1pjx&b.68.6.1  & 314& 307&0.147&0.789&0.884\\
1pq7&b.47.1.2  & 224& 214&0.187&0.609&0.795\\
1q6z&e.23.1.1  & 524& 523&0.105&0.769&0.843\\
1r2m&b.138.1.1 & 140& 135&0.156&0.704&0.843\\
1r6j&b.36.1.1  &  82&  70&0.043&0.363&0.484\\
1rb9&g.41.5.1  &  52&  48&0.146&0.304&0.888\\
1rtq&c.56.5.4  & 291& 291&0.076&0.573&0.684\\
1sfd&b.6.1.1   & 210& 210&0.148&0.674&0.626\\
1ssx&b.47.1.1  & 198& 174&0.115&0.734&0.793\\
1tg0&b.34.2.1  &  66&  56&0.161&0.792&0.837\\
1tqg&a.24.10.3 & 105&  71&0.014&0.405&0.575\\
1tt8&d.190.1.1 & 164& 164&0.079&0.657&0.678\\
1u2h&b.1.1.4   &  96&  96&0.115& 0.53&0.752\\
1ucs&b.85.1.1  &  64&  64&0.031&0.648& 0.67\\
1ufy&d.79.1.2  & 121& 118&0.093&0.769&0.784\\
1ug6&c.1.8.4   & 426& 426&0.103&0.738&0.833\\
1unq&b.55.1.1  & 117& 117&0.111&0.653&0.841\\
1us0&c.1.7.1   & 313& 259& 0.12&0.676&0.709\\
1v0l&c.1.8.3   & 302& 292&0.086&0.718&0.813\\
1v6p&g.7.1.1   & 124& 108&0.102&0.528&0.691\\
1vbw&d.40.1.1  &  68&  62&0.081&0.698&0.766\\
1vyr&c.1.4.1   & 363& 361&0.119&0.642& 0.79\\
1vyy&b.115.1.1 & 113&  99&0.121&0.648&0.714\\
1w0n&b.18.1.10 & 120& 116&0.121&0.495&  0.6\\
1x6z&d.24.1.1  & 119& 119&0.092&0.787&0.693\\
1x8q&b.60.1.1  & 184& 162&0.111&0.597&0.676\\
1xg0&d.184.1.1 & 494& 470& 0.17&0.597&0.666\\
1xmk&a.4.5.19  &  79&  75&0.093&0.584&0.846\\
1y55&b.61.1.1  & 240& 229&0.109&0.633&0.721\\
1ylj&e.3.1.1   & 263& 253&0.079&0.748&0.668\\
1zk4&c.2.1.2   & 251& 225&0.084&0.747&  0.8\\
1zzk&d.51.1.1  &  80&  78&0.103&0.723&0.701\\
2bt9&b.24.1.1  & 266& 262&0.164&0.685&0.786\\
2bw4&b.6.1.3   & 334& 292&0.103&0.831&0.901\\
2cws&b.29.1.18 & 227& 219&0.132&0.696&0.766\\
2f01&b.61.1.1  & 241& 238&0.151&0.612& 0.74\\
2fdn&d.58.1.1  &  55&  48&0.146&0.607&0.702\\
2pvb&a.39.1.4  & 107&  96&0.063&0.452&0.557\\
3lzt&d.2.1.2   & 129& 126&0.087&0.415&0.688\\
7a3h&c.1.8.3   & 300& 295&0.085&0.675&0.864\\

   \toprule
    \end{longtable}